\documentclass[
	twocolumn,
]{aastex631} %% apj

\usepackage{graphicx}% Include figure files
\usepackage{dcolumn}% Align table columns on decimal point
\usepackage{bm}% bold math
\usepackage{color}
\usepackage{xspace}
\usepackage{ulem}
\newcommand\raa{RAA}

\begin{document}
\title{The calibrations of DAMPE $\gamma$-ray effective area}

\newcommand{\affpmo}{Key Laboratory of Dark Matter and Space Astronomy, Purple Mountain Observatory, Chinese Academy of Sciences, Nanjing 210023, China}
\newcommand{\affustc}{School of Astronomy and Space Science, University of Science and Technology of China, Hefei, Anhui 230026, China}

\author{Zhao-Qiang~Shen}
\affiliation{\affpmo}

\author{Wen-Hao~Li}
\affiliation{\affpmo}
\affiliation{\affustc}

\author{Kai-Kai~Duan}
\affiliation{\affpmo}

\author{Wei~Jiang}
\affiliation{\affpmo}

\author{Zun-Lei~Xu}
\affiliation{\affpmo}
\affiliation{\affustc}

\author{Chuan~Yue}
\affiliation{\affpmo}
\affiliation{\affustc}

\author{Xiang~Li}
\altaffiliation{Corresponding author. Email: xiangli@pmo.ac.cn}
\affiliation{\affpmo}
\affiliation{\affustc}

%%%%%%%%%%%%%%%%%%%%%%%%%%%%% ABSTRACT %%%%%%%%%%%%%%%%%%%%%%%%%%%%%%%
\begin{abstract}
  The DArk Matter Particle Explorer (DAMPE) is a cosmic-ray detector as well as a pair-converting $\gamma$-ray telescope.
  The effective area, reflecting the geometrical cross-section area, the $\gamma$-ray conversion probability and the photon selection efficiency, is important in the $\gamma$-ray analyses.
  In the work, we find a significant time variation in the effective area, as large as $\sim -4\%/{\rm yr}$ at 2~GeV for the high-energy trigger.
  We derive the data-based correction factors to the effective areas and apply corrections to both the effective areas and the exposure maps.
  The calibrated exposure can be $\sim 12\%$ smaller than the Monte Carlo one on average at 2~GeV.
  The calibration is further verified using the observation of the Vela pulsar, showing the spectral parameters with the correction are more consistent with those in the \emph{Fermi}-LAT catalog than the ones without correction.
  All the corrections are now implemented in the latest version of the DAMPE $\gamma$-ray analysis toolkit {\sc DmpST}.
\end{abstract}

% \maketitle

\section{Introduction}\label{sect:intro}

%% instrument
Like EGRET~\citep{EGRET}, AGILE~\citep{AGILE} and \emph{Fermi}-LAT~\citep{Fermi2009,Fermi2021},
the DArk Matter Particle Explorer (DAMPE) is a pair-converting $\gamma$-ray telescope ~\citep{Chang2014}. With its good charge resolution, DAMPE is also a cosmic-ray detector that can measure the charged cosmic rays in a wide energy range~\citep{Chang2017}.
From top to bottom, DAMPE is comprised of four sub-detectors: the Plastic Scintillator strip Detector (PSD), the Silicon-Tungsten tracKer-converter (STK), the BGO imaging calorimeter (BGO), and the NeUtron Detector (NUD).
The PSD measures the charge of the incident particles and works as an anti-coincidence detector in $\gamma$-ray observations.
The STK converts the photons into secondary particles and measures their subsequent trajectories.
The BGO measures the deposited energy and images the shower profiles.
The NUD further enhances the electron/proton separation capability in the high-energy range~\citep{Chang2014,Chang2017,He:2023svm}.
%%%% current status
Since its launch on 17 December 2015, DAMPE has been performing stably for over eight years and collecting about two billion cosmic-ray (CR) events and about 40 thousand photons each year.
Thanks to the large CR effective area and the excellent charge and energy resolution, DAMPE accurately measures the spectra of various cosmic-ray species
and sets stringent constraints on the exotic particles~\citep[e.g.][]{DAMPE2017e,DAMPE2022BC,DAMPE2022fcp,DAMPE2022line,Cheng2023,Yang:2024jtp}.%,Fan:2024rcr}.

%% on calibrations
Studying the $\gamma$-ray emission from Galactic and extragalactic sources is one of the key science objectives of DAMPE.
To obtain accurate spectra of sources, the instrumental responses should be consistent with the actual ones.
Therefore, systematic calibrations of the instrumental responses are necessary.
Currently, the on-orbit calibrations in the sub-detector level have been conducted~\citep{Ambrosi2019,Huang:2020skz}.
The alignments among the sub-detectors~\citep{Cui2023,Ma2019} and between the payload and the satellite~\citep{Jiang2020} have also been carried out.
However, due to the complexity of the instrument, a good understanding of the instrument configuration does not automatically ensure the simulated responses are consistent with the actual ones.
So direct calibrations with observational data are important as well.

%% main purpose and structure of the paper
The $\gamma$-ray responses of DAMPE are parameterized with the three instrumental response functions (IRFs): the effective area, the point-spread function (PSF), and the energy dispersion functions~\citep{Duan2019}.
The PSF is calibrated using the photons from the active galaxy nuclei and pulsars in the accompanying paper~\citep{Duan2024}.
Recently, the attenuation lengths of the BGO crystals are found to be decreasing at the rate of $2.7\%/\rm yr$ on average, due to the continuous radiation bombardment, mainly from the trapped electrons and protons in the geomagnetic field~\citep{Liu:2023ibg}. %% 4 mm/mo * 12 mo/yr / (1800 mm)
The accumulated radiation damage increases the high-energy trigger thresholds by $0.9\%/\rm yr$ on average for different BGO bars, which naturally leads to the time-dependent variation of the electron effective area~\citep{Li2023}.
In this work, we adopt a different approach from~\citet{Li2023} to calibrate the effective area, paying special attention to that of photons.
We will not only study the variations for events with different trigger types, but also analyze the dependence of the variation rates to the energy and inclination angle.
This paper is structured as follows:
The time variations of the effective areas are derived in Sec.~\ref{sect:derivation} and the influence of the inclination angle is evaluated in Sec.~\ref{sect:theta_correction}.
The parameterized corrections are then proposed in Sec.~\ref{sect:correction} and verified using the Vela pulsar in Sec.~\ref{sect:verify}.
This work is summarized in Sec.~\ref{sect:summary}.

%%% Sec. 2
\section{Time variation of the effective area}\label{sect:derivation}
The effective area is the actual area which the instrument collects photons.
It is the product of the geometrical cross-section area, the $\gamma$-ray conversion probability, and the photon selection efficiency~\citep{Duan2019}.
The selection efficiency depends on the thresholds of the trigger selection, the electron/proton discrimination conditions, and the algorithm of the track selection and the charged particle rejection~\citep{Xu2018,Xu2022}.
The DAMPE $\gamma$-ray effective area $A_{{\rm eff},s}(E,\hat{v})$ is the function of the primary energy $E$, the incident direction $\hat{v}$ and the event type ({\tt evtype}) $s$.
The Low-Energy trigger (LET) and the High-Energy trigger (HET) are the two types of trigger logics adopted in $\gamma$-ray observations, which aim for the photon events with primary energy $\gtrsim 1~\rm GeV$ and $\gtrsim 5~\rm GeV$, respectively~\citep{Zhang2019}.
These logics are assigned by comparing the electronics signal amplitudes of the crystals to the trigger thresholds in the top four BGO layers~\citep{Zhang2019,Li2023}.
The $\gamma$-ray data are divided into two event types, which correspond to two combinational trigger logics~\citep{Duan2019}:
The {\tt evtype=H} events and IRFs satisfy the trigger logic of $\rm HET$, denoted as the HET events and HET IRFs;
the {\tt evtype=L} events and IRFs satisfy the $\rm LET \& \overline{HET}$ logic, denoted as the HET vetoed LET (HvLET) events and HvLET IRFs.
The HvLET events are pre-scaled by a factor of $\approx 1/8$ ($\approx 1/64$) on the satellite in the low (high) geographic latitude region~\citep{Chang2017,Zhang2019}.
The current effective areas are derived with the Monte Carlo (MC) simulation given a fixed trigger thresholds~\citep{Duan2019,Xu2018,Chang2017}, which is denoted as $A_{\rm eff}^{\rm MC}$.

In this section, we use the observed photons to directly calibrate the time variations of the $\gamma$-ray effective areas.
Considering that the all-sky $\gamma$-ray flux is generally stable, the ratio of the observed photon counts to the calculated counts using MC IRFs can reflect the variation of the effective areas.
Because the HET and LET triggers have different thresholds, we discuss them separately in the two subsections below.
In the calibration, we choose the {\tt v6.0.3} photon data set~\citep{Xu2018,Jiang2020,Shen2023ICRC} collected from 2016 January 1 to 2023 July 1, and exclude the events during the intervals when the telescope is in the South Atlantic Anomaly (SAA) region or is strongly affected by the solar flares.

\subsection{High-energy-trigger effective area}\label{sect:derivation:HET}
\begin{figure}[!tb]
  \centering
  \includegraphics[width=\columnwidth]{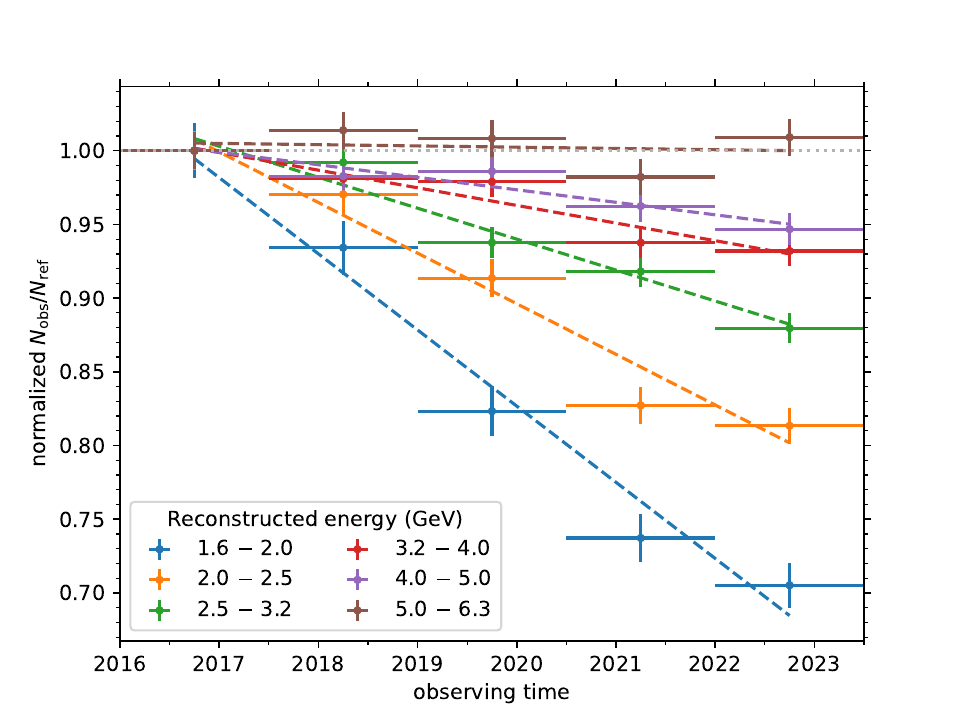}
  \caption{
    The ratios of the observed High-Energy Trigger (HET) photon counts to the reference counts in different time bins.
    The ratios are normalized to 1.0 at the first bin.
    Different colors represent different energy bins.
    The dashed lines present the linear models that fit the variations best.
  }\label{fig:het_ratio_E}
\end{figure}

We first evaluate the variation of the HET effective area by comparing the observed all-sky counts with the reference ones in different time bins.
The reference counts are based on the best-fit $\gamma$-ray model derived from the entire data set.
The HET photon events between 2~GeV to 500~GeV are selected and binned according to the HEALPix projection~\citep{Healpix2005} with {\tt NSIDE=64}.
The 2-deg regions around the sources with TS values\footnote{
  The TS value is defined as ${\rm TS} \equiv -2\ln (\hat{L}_{\rm null}/\hat{L}_{\rm sig})$, where $\hat{L}_{\rm null}$ and $\hat{L}_{\rm sig}$ are the likelihood values of the best-fit null (exclude the test source) and alternative (contain the test source) models, respectively~\citep{Cash1979,Mattox1996}.
} larger than 100 in the 7.5-yr DAMPE point source catalog~\citep{Duan2023ICRC,DAMPECatalog} are masked to reduce the potential influence from the variable sources.
The data are fitted with the \emph{Fermi}-LAT Galactic diffuse emission model {\tt gll\_iem\_v07}~\citep{FermiGDE2016,4FGL2022},\footnote{\url{https://fermi.gsfc.nasa.gov/ssc/data/access/lat/BackgroundModels.html}} the isotropic template model with a power-law spectral shape~\citep{Ackermann2015_IGRB}, and the stacked DAMPE point source template~\citep{Duan2023ICRC,DAMPECatalog}.\footnote{
  No significant change occurs if we switch to the 14-yr \emph{Fermi}-LAT point source catalog {\tt gll\_psc\_v32}~\citep{4FGL2022}.
}

%%%% ratio-E relation for HET
We first evaluate the energy-dependent component in the variation of the effective area.
In Fig.~\ref{fig:het_ratio_E}, we present the ratios between the observed counts and reference counts in different reconstructed energy bins as a function of observing time.
The ratios should be independent of the observation mode of the telescope, since the exposure time in the numerator and denominator is the same and is canceled out in the ratio.
The width of the time bin is around 1.5 years.
We also extend the derivation to a lower energy bin of $1.6~{\rm GeV} - 2~{\rm GeV}$.
As is clearly shown in the figure, the ratios decrease with time and the trend is more significant for lower energy.
Since the all-sky $\gamma$-ray flux is generally stable, the variation of the ratios should come from the change in the effective area.

\begin{figure}[!tb]
  \centering
  \includegraphics[width=\columnwidth]{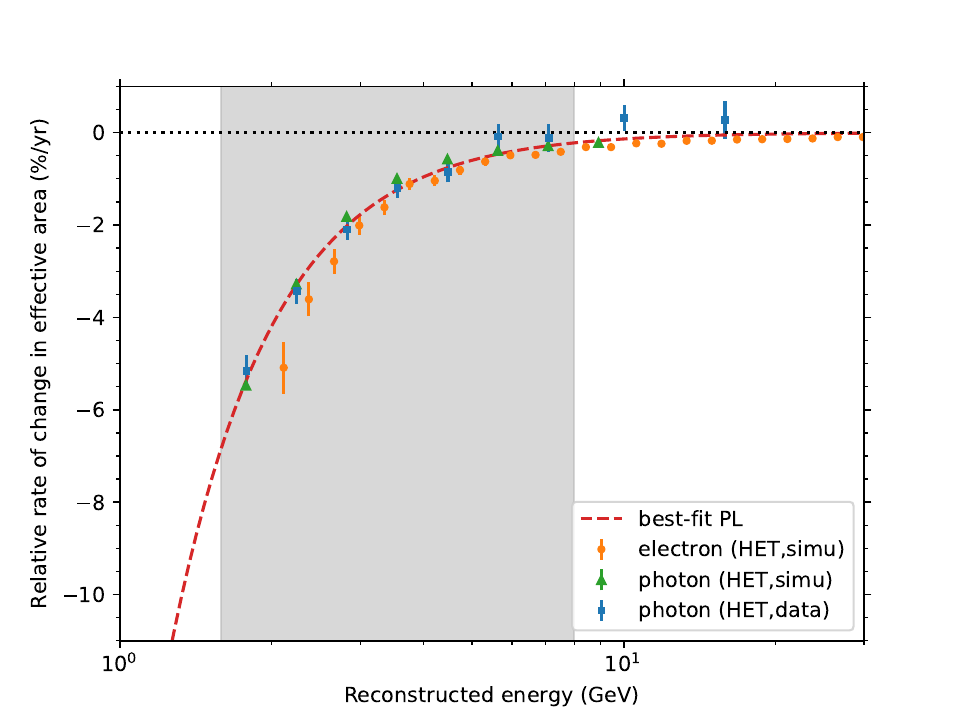}
  \caption{
    The relative rate of change in the HET photon effective area (blue points) concerning the reconstructed energy.
    The red dashed line shows the power-law (PL) model fitted with the points whose reconstructed energies are $\lesssim 8~\rm GeV$ (shaded region).
    The orange and green points are the expected change rates of HET efficiencies of electron~\citep{Li2023} and photon in simulation, respectively.
  }\label{fig:het_ratio_rate_E}
\end{figure}

In Fig.~\ref{fig:het_ratio_rate_E}, we show the derived relative rates of change in the HET effective area with the blue points.
The relative decrease rate of photon HET effective area is about $4\%/{\rm yr}$ at 2~GeV and is $\lesssim 1\%/{\rm yr}$
at the reconstructed energy higher than $\sim 5~\rm GeV$.
We also depict the change rates of the electron efficiency presented in~\citep{Li2023} with orange points, which are derived from the simulation considering the increasing trigger thresholds of the BGO bars.
The variation rate of photon effective area shares a similar trend as that of the electron.
So the photon effective area variation also originates from the aging of BGO crystals and associate electronics.
It is noted that the change rates of the photons are slightly smaller than those of the electrons with the same reconstructed energy.
Since the photons usually initiate showers later in the STK than the electrons with same primary energy,
the showers lose less energy on their way to the calorimeter and deposit more energy into the first four BGO layers,
leading to a higher probability to trigger the HET logic for $\gamma$ rays.

We fit the relative rates of change with respect to the reconstructed energy $E$ in the energy band $\lesssim 8~\rm GeV$ with the power-law (PL) model
\begin{equation}\label{eq:f}
  -f(E) = A \times (E/{\rm 3~GeV})^\gamma.
\end{equation}
The best-fit model is shown with the red dashed line in Fig.~\ref{fig:het_ratio_rate_E}.
The optimized parameters are $A=0.0177\pm0.0012~\rm yr^{-1}$ and $\gamma=-2.12\pm0.17$.
The $\chi^2$ is 3.32 for 5 degrees of freedom (dof), so the power-law model is good enough to represent the points.

%%%% LET
\subsection{Low-energy-trigger effective area}
As introduced previously, the HvLET photon events fulfil the LET logic but not the HET logic, so this data set is mainly comprised of the events with the primary energies between $\sim 1~\rm GeV$ and $\sim 5~\rm GeV$.

\begin{figure}[!tb]
  \centering
  \includegraphics[width=\columnwidth]{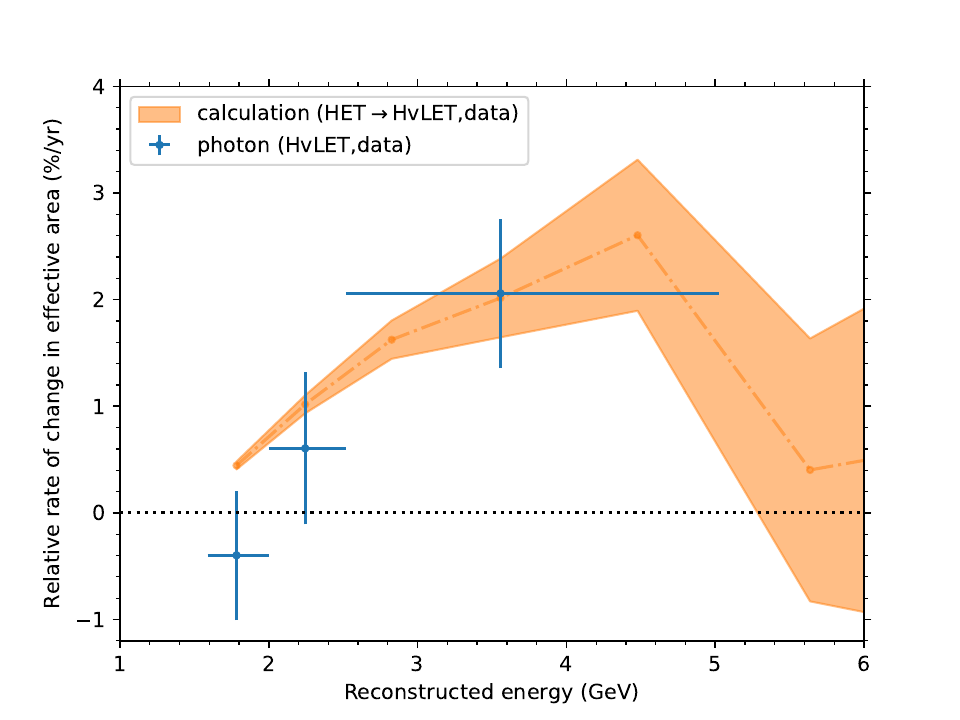}
  \caption{
    The relative rate of change in the HvLET photon effective area (trigger logic $\rm LET \& \overline{HET}$) with respect to the reconstructed energy.
    The blue points are derived from the observed events.
    Shown in orange is the expected contribution of the HET photons which do not satisfy the HET logic due to the increasing HET threshold energy.
    The dot-dashed line and the color band correspond to the best-fit values and the $1\sigma$ uncertainties of the relative decrease rates of the HET effective area, respectively.
  }\label{fig:let_ratio_rate_E}
\end{figure}

In Fig.~\ref{fig:let_ratio_rate_E}, we present the relative rate of change in the HvLET effective area in the energy range between 1.6~GeV and 5~GeV based on the same method detailed in Sec.~\ref{sect:derivation:HET}.
Because of the low statistics, we can not reliably derive the results in higher energies.
As shown in the figure, the effective area gradually increases with energy and reaches $2\%/{\rm yr}$ at 3~GeV.

Theoretically speaking, the variation of the effective area comes from two components:
the decrease of the LET effective area because of the increasing LET threshold energy,
and the increase from the photons that should satisfy the HET logic but do not due to the increasing HET threshold energy (`failed' HET photon events).
The latter one, as shown with the orange band in Fig.~\ref{fig:let_ratio_rate_E}, can be estimated by multiplying the HET photon counts considering the pre-scale factors by the relative decrease rate of the HET effective area.
The dot-dashed line and the color band correspond to the best-fit values and the $1\sigma$ statistic uncertainties of the decrease rates, respectively.
The trend of this component is shaped by the decrease of `failed' HET photon counts and the decrease of HvLET effective area as the energy grows.
Compared with the change rate of HET effective area, it shrinks at lower energy but enhances at higher energy.
The change rate increases below $\sim 5~\rm GeV$, reaches a peak of $\sim 2.5\%/{\rm yr}$, and decreases above $\sim 5~\rm GeV$.

As illustrated in the figure, the variation of the HvLET effective area can be well explained by the `failed' HET events in the energy range $\gtrsim 2~\rm GeV$.
On the other hand, the increasing LET thresholds seem to dominate the variations of the HvLET effective area below $\sim 2~\rm GeV$.
Because only the photons with reconstructed energies higher than $2~\rm GeV$ are adopted in the standard $\gamma$-ray analyses, we do not account for the variation of the LET effective area.

%%% Sec. 3.
\section{Influence of inclination angle to the variation rate}\label{sect:theta_correction}

%%%% ratio-ctheta relation for HET
\begin{figure}[!tb]
  \centering
  \includegraphics[width=\columnwidth]{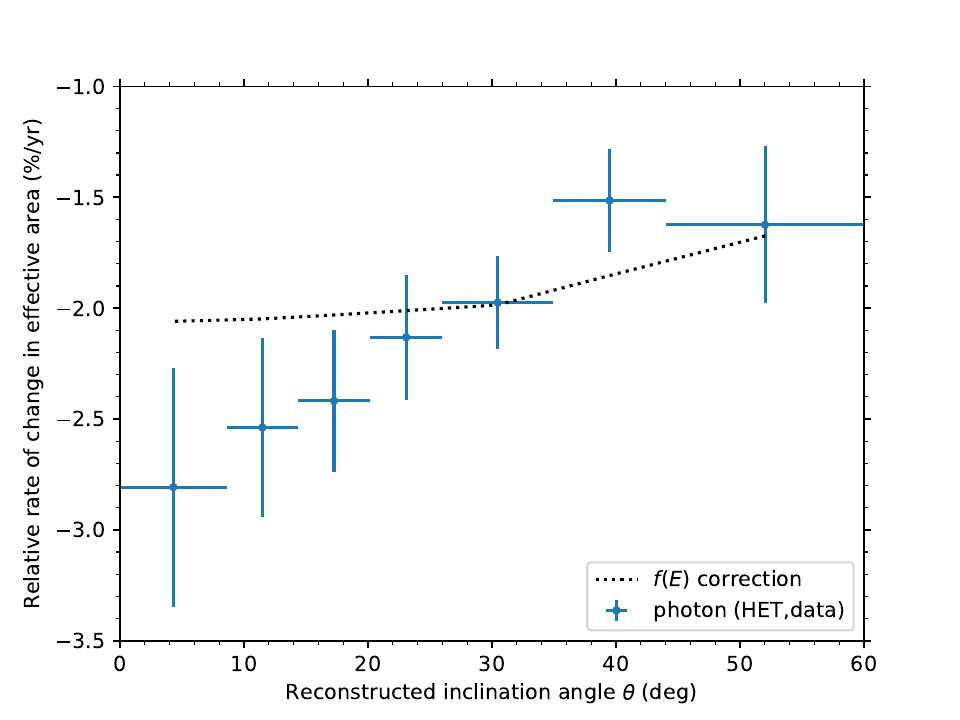}
  \includegraphics[width=\columnwidth]{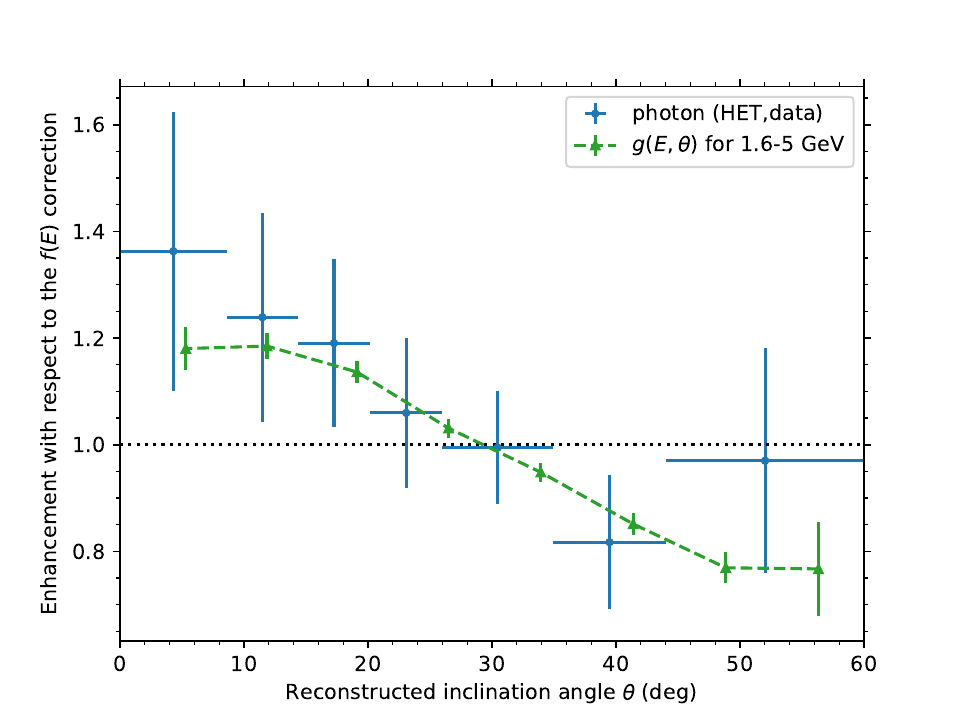}
  \caption{
    \emph{Upper panel:}
    The relative rate of change in HET photon effective area with respect to the reconstructed inclination angle $\theta$ (blue points) derived with the data below 5~GeV.
    The black dotted line is the rate of change only corrected with the energy-dependent factor $f(E)$.
    \emph{Lower panel:}
    The relative rate of change in the HET effective area divided by the relative rate from the factor $f(E)$.
    The green points and dashed line represent the enhancement correction factor $g(E,\theta)$ derived from the simulated photon data between 1.6~GeV and 5~GeV.
  }\label{fig:het_ratio_ctheta}
\end{figure}

\begin{figure}[!tb]
  \centering
  \includegraphics[width=0.775\columnwidth]{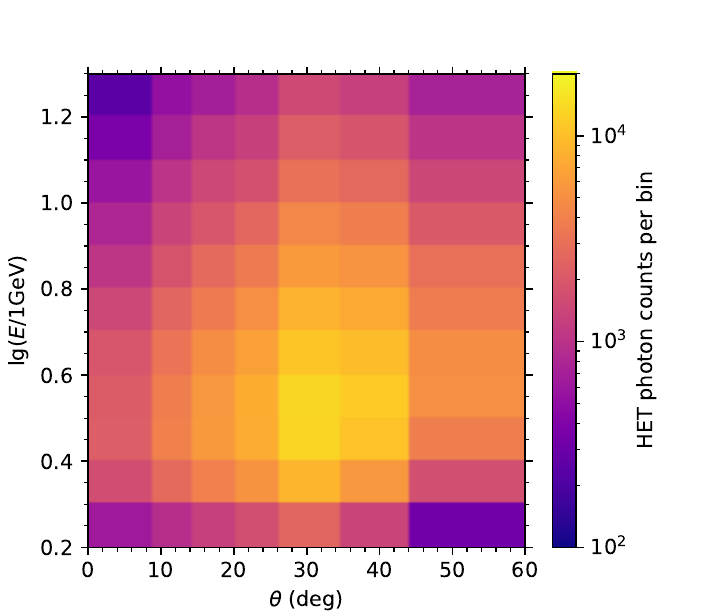}
  \caption{
    The counts of the HET photon data binned according to the reconstructed energy and inclination angle.
    The $E$ bins are logarithmic from 1.6~GeV to 20~GeV, while the $\theta$ bins match those in Fig.~\ref{fig:het_ratio_ctheta}.
  }\label{fig:het_counts_E}
\end{figure}

Besides the energy-dependent factor, an inclination-angle dependency ($\theta$ dependency) may also exist in the variation rate of the effective area.
If there was such a dependency, the change rates of the exposure would be different in various sky positions, which may bias the $\gamma$-ray analyses.
In this section, we use the flight data to derive the additional $\theta$-dependent variation on top of the energy-dependent factor $f(E)$.
Then the MC simulations are made to calculate the additional correction factor $g(E,\theta)$.

\subsection{The variation rate from the flight data}
We split the HET photon data below 5~GeV according to the reconstructed inclination angle $\theta$, and derive the relative rates of change in different $\theta$ bins as shown in the upper panel of Fig.~\ref{fig:het_ratio_ctheta}.
The black dotted line presents the relative rates corrected with the energy-dependent factor $f(E)$, which is $ \sim -2\%/{\rm yr}$ in the analyzed energy range.
The weak $\theta$ dependence of the $f(E)$ correction comes from the different energy distributions of events in different $\theta$ bins as illustrated in Fig.~\ref{fig:het_counts_E}: the change rate in the larger inclination angle is smaller because of fewer low-energy photons.
We further divide the observed rate by the correction from the energy-dependent factor $f(E)$ as depicted in the lower panel of Fig.~\ref{fig:het_ratio_ctheta}.
A slight $\theta$ dependence is visible in the change rate of effective area, but the energy-dependent correction alone can also explain the flight data quite well since $\chi^2/{\rm dof} = 7.2/7$.

\subsection{The variation rate from the simulation}
To further quantify the $\theta$ dependence of the effective area change rate, we take advantage of the MC simulation.
The simulation photon data are generated using {\sc Geant4}~\citep{GEANT4:2002zbu,Allison:2006ve,GEANT4} based on the geometric model consisting of both the payload and the satellite platform.
We generate photon events in the detector, do particle transportation simulation, and convert the raw hits into digital signals~\citep{Xu2018,Jiang:2020nph}.
After the digitization, the trigger thresholds of the top four BGO layers at a specific time are calculated based on the increase rates in~\citet{Li2023} and applied to the MC data.
The reconstructions are then performed.% to make simulated photons.
We carried out three simulations with the trigger thresholds at 0~days, 1500~days, and 3000~days after the launch.

Taking advantage of the simulated data, we make the effective areas in the bins of primary energy $E$ and inclination angle $\theta$~\citep{Duan2019}.
We denote these three sets of the simulated effective areas as $A_{\rm eff}^{{\rm MC},i}\equiv A_{\rm eff}^{\rm MC}(E,\theta,t_i)$, where $t_0$, $t_1$ and $t_2$ refer to 0~days, 1500~days and 3000~days after the launch, respectively.
The time evolution of the HET effective areas is illustrated in Fig.~\ref{fig:het_simuaff_evo} in the appendix.
The same as that in the flight data (Fig.~\ref{fig:het_ratio_E}), the evolution of the MC effective areas can also be well modeled with linear functions.
The acceptance can also be calculated using $\varepsilon(E,t_i) \equiv \int A_{\rm eff}^{{\rm MC},i}(E,\theta){\rm d}\Omega$.
In Fig.~\ref{fig:het_ratio_rate_E}, we present the relative change rates of the acceptance with green points, which are derived by fitting the relative acceptances $\varepsilon(E,t_i)/\varepsilon(E,t_0)$ with linear evolving functions.
The low-energy acceptance changes faster than the high-energy one because the photons with higher primary energy are less likely influenced by the growing trigger thresholds.
At very high energy ($\gtrsim 30~\rm GeV$), the change rate would be negligible.
In the lower panel of Fig.~\ref{fig:het_ratio_ctheta}, we draw the ratio of the relative change rate of the simulated effective area, integrated from 1.6~GeV to 5~GeV, to the relative rate of the acceptance (green points).
Both results show the simulations with the changed trigger thresholds consistent with the on-orbit $\gamma$-ray data, proving the reliability of the bottom-up calibration conducted by~\citet{Li2023}.

\begin{figure}[!tb]
  \centering
  \includegraphics[width=0.9\columnwidth]{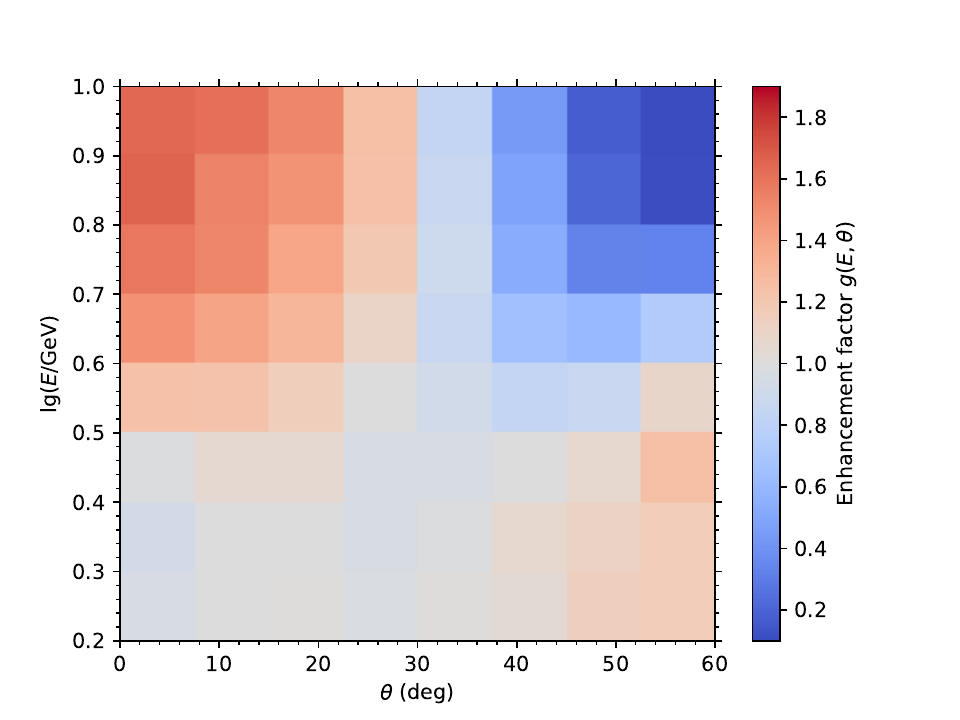}
  \caption{
    The enhancement of the variation rates on top of the simple energy-dependent correction.
    It is derived by dividing the relative MC effective area change rates by the relative acceptance change rate.
    The $\theta$ dependence is weak for the primary energy between 1.6~GeV and $\sim 3$~GeV.
  }\label{fig:het_gEth_mat}
\end{figure}

To calculate the enhancement of the variation rates on top of the simple energy-dependent correction, we divide the relative change rates of the MC effective area $\dot{A}_{\rm eff}^{{\rm MC}}(E,\theta,t)/A_{\rm eff}^{{\rm MC},0}$ by the change rate of the acceptance $\dot{\varepsilon}(E,t)/\varepsilon(E,t_0)$.
The results are shown in Fig.~\ref{fig:het_gEth_mat}.
The $\theta$ dependence is weak for the primary energy between 1.6~GeV and $3$~GeV but increases in the higher energy range.
For low-energy photons with larger inclination angles, it is a bit harder for secondary particles to reach the fourth BGO layer to meet the HET logic, therefore the variation rate increases slightly for large inclination angles.
For photons with higher energies, those injected with a larger inclination angle will deposit more energy in the top four calorimeter layers and therefore are less influenced by the increasing HET threshold.

We make an interpolation function $g(E,\theta)$ based on the enhancement factor of the variation rates in Fig.~\ref{fig:het_gEth_mat} to facilitate the calibration of the effective area.

\begin{figure*}[!tb]
  \centering
  \includegraphics[width=0.95\columnwidth]{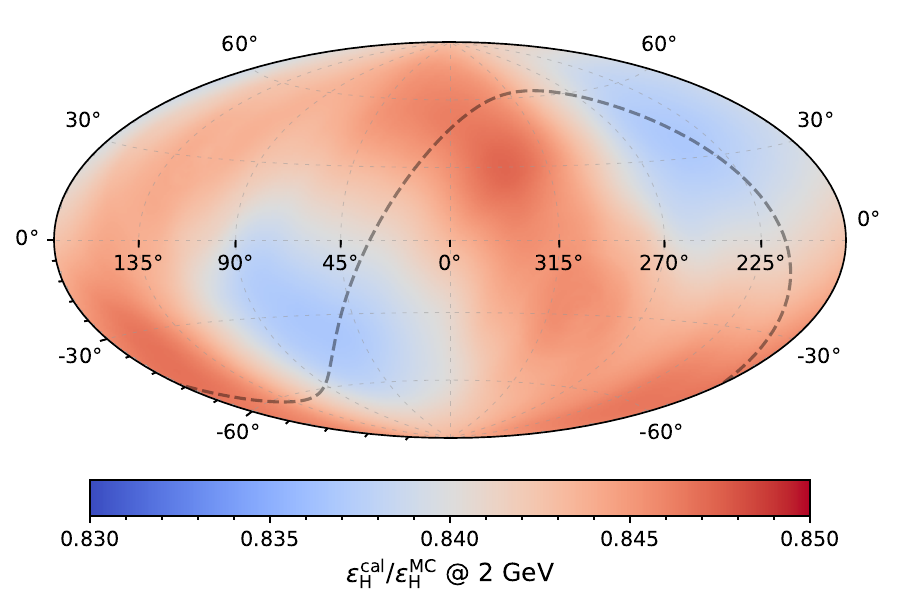}
  \includegraphics[width=0.95\columnwidth]{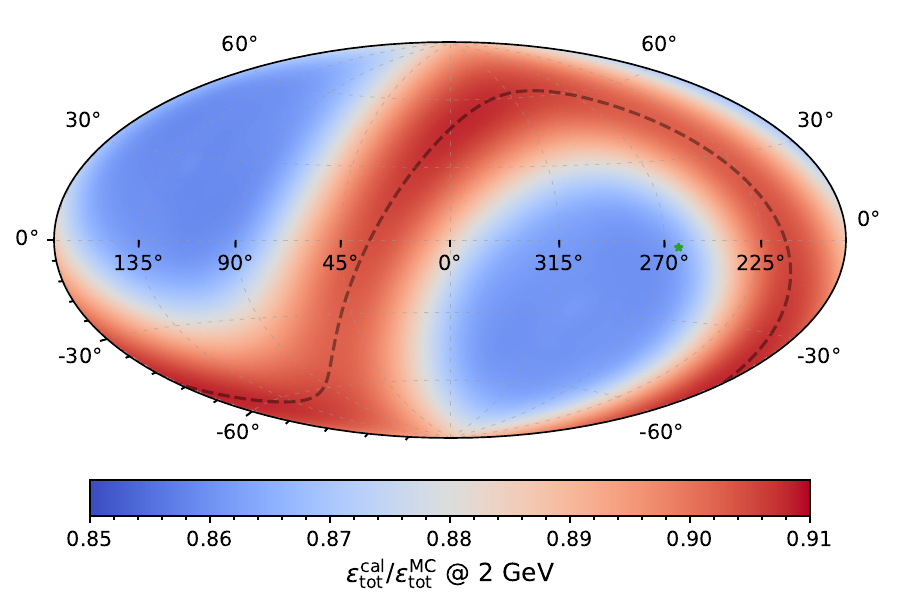}
  \caption{
    The ratio of the exposure map calibrated with the $f(E) g(E,\theta)$ correction to the MC one at the primary energy of 2~GeV.
    The left and right panels correspond to the HET exposure $\varepsilon_{\rm H}$ and total exposure $\varepsilon_{\rm tot} \equiv \varepsilon_{\rm H}+\varepsilon_{\rm L}$, respectively.
    The map is binned according to the Aitoff projection in the Galactic coordinate.
    The black dashed line represents the celestial equator.
    The green star in the right panel shows the location of the Vela pulsar.
  }\label{fig:exposure_ratio}
\end{figure*}

%%% Sec. 3
\section{Parameterized correction}\label{sect:correction}
We adopt the following parameterized correction to the HET effective area ({\tt evtype=H})
\begin{equation}
  A_{\rm eff,H}^{\rm cal}(E,\theta,\phi, t) = A_{\rm eff,H}^{\rm MC}(E,\theta,\phi) \times \left[ 1 - f(E)g(E,\theta) \Delta t \right],
\end{equation}
where $E$ is the primary energy, $\theta$ is the incident inclination angle, $\phi$ is the incident azimuth angle, and the subscript `H' denotes {\tt evtype=H}.
We factorize the change rate of the effective area into two correction factors:
one is the primary factor $f(E)$ derived from the flight data as defined in Eq.(\ref{eq:f});\footnote{
  Even though $f(E)$ should be the function of the reconstructed energy, using primary energy instead will not significantly change the results because of the small energy dispersion comparing to the $E$ bins adopted in the derivation.
}
the other is the secondary factor $g(E, \theta)$ based on the MC simulations with the trigger thresholds changed as detailed in Sec.~\ref{sect:theta_correction}.
This form of correction takes advantage of both the direct information from the flight data and the fine tuning from the simulations with relatively large statistics.
We do not apply the correction about $\phi$ due to the limited data.
$\Delta t \equiv t - t_{\rm ref}$ and $t_{\rm ref}$ is the time when $A_{\rm eff}^{\rm MC}$ is consistent with the observation.
Considering that lots of the DAMPE on-orbit calibration are based on the data collected between 2015 December 24 and 2017 April 1~\citep{Ambrosi2019}, we choose 2016 July 1 as $t_{\rm ref}$ which is approximately at the middle of the time interval of calibration, i.e. $t_{\rm ref}=110332801$ in DAMPE Mission Elapsed Time.

The modified HET exposure for the source in the sky $\hat{p}$ and at the photon energy $E$ is
\begin{eqnarray}
  \varepsilon^{\rm cal}_{\rm H}(E,\hat{p}) 
  &\equiv& \int A_{\rm eff,H}^{\rm cal}(E,\hat{v}(\hat{p},t), t) \, {\rm d}t_{\rm lt} \nonumber \\
  &=& \int A_{\rm eff,H}^{\rm MC}(E,\hat{v})\, {\rm d}t_{\rm lt}  \nonumber \\
  &&- f(E) \int A_{\rm eff,H}^{\rm MC}(E,\hat{v})\,g(E,\theta)\,\Delta t \, {\rm d}t_{\rm lt} \nonumber \\ 
  &=& \varepsilon^{\rm MC}_{\rm H}(E,\hat{p}) - f(E) \, \varepsilon^{\rm corr}_{\rm H}(E,\hat{p}),
\end{eqnarray}
where ${\rm d}t_{\rm lt}(t)$ is the live time in the observing time interval from $t$ to $t + {\rm d}t$.
The correction is $\varepsilon^{\rm corr}_{\rm H}(E,\hat{p})=\int A_{\rm eff,H}^{\rm MC}(E,\hat{v}) \, g(E,\theta) \, \Delta t \, {\rm d}t_{\rm lt}$.

We also correct the HvLET effective area ({\tt evtype=L}) with
\begin{eqnarray}
  A_{\rm eff,L}^{\rm cal}(E,\theta,\phi, t) &=& 
  A_{\rm eff,L}^{\rm MC}(E,\theta,\phi) \nonumber \\
  &+& A_{\rm eff,H}^{\rm MC}(E,\theta,\phi) f(E)g(E,\theta) \Delta t,
\end{eqnarray}
where the second term on the right comes from the `failed' HET events.
The modified exposure therefore becomes
\begin{eqnarray}
  \varepsilon^{\rm cal}_{\rm L}(E,\hat{p}) 
  &\equiv& \int \eta(\phi_{\rm geo}(t)) \, A_{\rm eff,L}^{\rm cal}(E,\hat{v}(\hat{p},t), t) \, {\rm d}t_{\rm lt} \nonumber \\
  &=& \int \eta(\phi_{\rm geo}(t)) \, A_{\rm eff,L}^{\rm MC}(E,\hat{v}) \, {\rm d}t_{\rm lt} \nonumber \\
  &+& f(E) \int \eta(\phi_{\rm geo}(t)) \, A_{\rm eff,H}^{\rm MC}(E,\hat{v}) \,g(E,\theta) \,\Delta t \, {\rm d}t_{\rm lt}   \nonumber \\
  &=& \varepsilon^{\rm MC}_{\rm L}(E,\hat{p}) + f(E)\, \varepsilon^{\rm corr}_{\rm L}(E,\hat{p}),
\end{eqnarray}
where $\eta(\phi_{\rm geo})$ is the pre-scale factor, $\phi_{\rm geo}$ is the geometrical latitude of the satellite, and the subscript `L' denotes {\tt evtype=L}.
$\varepsilon^{\rm corr}_{\rm L}(E,\hat{p})=\int A_{\rm eff,H}^{\rm MC}(E,\hat{v}) \, g(E,\theta) \, \eta(\phi_{\rm geo}(t))\, \Delta t\, {\rm d}t_{\rm lt}$.

Both the parameterized corrections, implemented in the latest version of the $\gamma$-ray analysis toolkit {\sc DmpST},\footnote{\url{https://dampe.nssdc.ac.cn/dampe/dampetools.php}}
are accounted for in the recent DAMPE diffuse $\gamma$-ray analyses~\citep{Shen2023ICRC}.
The correction factors will also be updated regularly along with the toolkit.

We compare the 7.5-yr exposure map before and after the correction.
For the HET exposure, the calibrated one is $\sim 16\%$ smaller than the original MC one on average at 2~GeV.
When the photon energy is higher than 8~GeV, the difference shrinks to $\lesssim 1\%$ and can be neglected in the analyses.
We show the ratio of the calibrated exposure to the MC one at the primary energy of 2~GeV in the left panel of Fig.~\ref{fig:exposure_ratio}.
Since DAMPE surveys the full-sky twice a year, a slightly decrease of effective area within each survey causes the exposure difference exists along the celestial longitude.
Thankfully, the morphology difference in the exposure maps is only $\sim 2\%$, and therefore it should not affect the structure study of diffuse emission when only the HET data are adopted.

In the right panel of Fig.~\ref{fig:exposure_ratio}, we also show the ratio of the total exposure maps before and after the calibration at 2~GeV.
The average calibrated exposure is $\sim 12\%$ smaller than the MC one.
The spatial difference is as large as $6\%$, which is caused by the different pre-scale factors of the HvLET events in different latitudes.

%%% Sec. 4
\section{Verification}\label{sect:verify}
\begin{figure}[!tb]
  \centering
  \includegraphics[width=0.9\columnwidth]{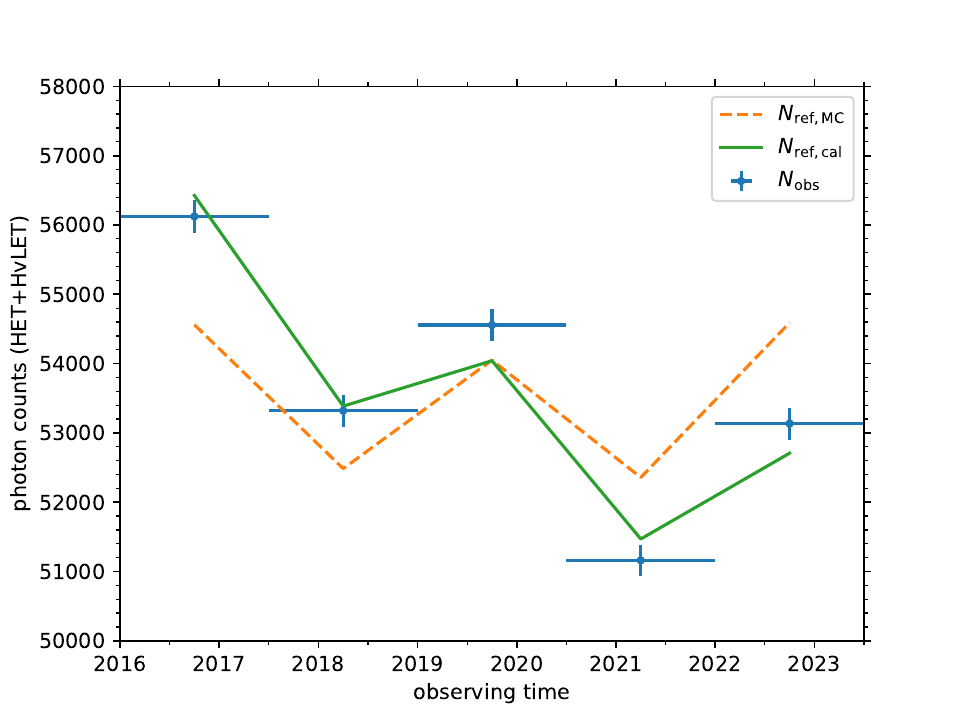}
  \caption{
    The variations of the observed counts and expected counts between 2~GeV and 20~GeV.
    The observed counts $N_{\rm obs}$ are marked with blue points.
    The expected counts with correction $N_{\rm ref,cal}$ and without correction $N_{\rm ref,MC}$ are shown with green solid line and orange dashed line, respectively.
    The expected counts are based on the best-fit models of the data set above 2~GeV from January 2016 to July 2023.
  }\label{fig:count_var}
\end{figure}

The variation of the effective area causes a mismatch between the observed counts and the expected counts accross various time bins.
In Fig.~\ref{fig:count_var}, we illustrate the variation of the observed counts $N_{\rm obs}$ and the predicted counts $N_{\rm ref}$ from 2~GeV to 20~GeV.
The predicted counts are based on the best-fit model of the entire data set above 2~GeV.
The $\gamma$-ray model remains the same as in Sec.~\ref{sect:derivation}.
The best-fit normalization of the Galactic diffuse emission {\tt gll\_iem\_v07} is $0.969\pm0.003$ for the model without correction; while the value is $1.008\pm0.003$ for the one with correction.
As shown in the figure, the counts without correction $N_{\rm ref,MC}$ (orange dashed line) are relatively stable, different from the gradually decreasing trend of the observed counts ($\chi^2=129.0$).
Conversely, the counts with correction $N_{\rm ref,cal}$ (green solid line) show a much better agreement with the observation ($\chi^2=11.9$).
Thus, the parameterized correction defined in Eq.(\ref{sect:correction}) effectively resolves the mismatch in counts variation.

The variation of the effective area also distorts the low-energy spectra of sources.
We adopt the Vela pulsar, the brightest source in the DAMPE point source catalog, to verify the parameterized correction to the effective area.
It is marked with a green star in the right panel of Fig.~\ref{fig:exposure_ratio}.
If the correction is valid, the derived spectral parameters should be more consistent with those in the \emph{Fermi}-LAT catalog comparing to the parameters without the calibration.

\begin{table}
  \centering
  \caption{\label{tab::verify}
      The natural logarithm of the largest likelihood values in the analyses of Vela pulsar given different types of effective area corrections.
      $\Delta \ln L \equiv \ln L - \ln L_{\rm MC}$.
      Details can be found in Sec.~\ref{sect:verify}.
  }
  \begin{tabular}{l|cc}
   \hline\hline
   model & $\ln L$ & $\Delta \ln L$ \\
   \hline
   no correction (MC $A_{\rm eff}$)    & $-12485.4$ & $\cdots$ \\
   $f(E)$ correction          & $-12460.0$ & $25.4$   \\
   $f(E)g(E,\theta)$ correction & $-12460.2$ & $25.2$   \\
   \hline\hline
  \end{tabular}
\end{table}

We select $2-200~\rm GeV$ photons within the $6^\circ \times 6^\circ$ rectangular region around the pulsar ($\alpha = 128 \fdg 837$, $\delta = -45 \fdg 1781$~\citep{4FGL2022}) from the HET and HvLET data sets.
Those photons are binned into $120 \times 120$ pixels and 16 logarithmically spaced energy bins.
The $\gamma$-ray emission model consists of three components: a point source representing the Vela pulsar, the Galactic diffuse emission {\tt gll\_iem\_v07}~\citep{FermiGDE2016,4FGL2022}, and the isotropic template.
For the Vela, we adopt the spectral parameters from the \emph{Fermi}-LAT 4FGL-DR4 point source catalog {\tt gll\_psc\_v32} \citep{4FGL2022} and keep them fixed in the fitting.
We also fix the normalization of the Galactic emission model to 1.0 and only optimize the prefactor and index of the isotropic component.

As shown in the Tab.\ref{tab::verify}, the models with the effective area corrections better fit the photon data, and the likelihood value differences $\Delta\ln L$ are larger than $\sim 25$.
It means that the calibrated exposure makes the spectral parameters in the \emph{Fermi}-LAT catalog more consistent with the DAMPE data.
However, 
the one corrected with $f(E)g(E,\theta)$ does not further improve the fitting compared to that corrected with $f(E)$.
Since the live-time-weighted inclination angle of the Vela is about $40^\circ$ for DAMPE and the enhancement $g(E,\theta)$ factor is very close to 1.0 (Fig.~\ref{fig:het_gEth_mat}), it can be reasonable that the additional correction only makes a slight difference to the analysis of Vela.

%%% Summary
\section{Summary}\label{sect:summary}
The aging of the sensitive units is a common issue for space-borne instruments due to the continuous radiation bombardment.
DAMPE is a cosmic-ray detector and $\gamma$-ray telescope, collecting data stably for more than eight years.
In this work, we show with the DAMPE $\gamma$-ray flight data that this effect will reduce the effective area around the threshold energy, which may potentially bias the spectral and morphological analyses.
We also develop a data-based method to calibrate the variation of the $\gamma$-ray effective area, which may also be valuable for other space telescopes.

A significant time variation of the HET effective area, around $-4\%/{\rm yr}$ at the reconstructed energy of 2~GeV, is found by comparing the observed photon counts with the prediction using the MC IRFs.
This variation can be attributed to the increasing HET thresholds induced by the radiation damage of the BGO calorimeter as shown in Fig.~\ref{fig:het_ratio_E}.
We derive the correction factors to the HET effective area as detailed in Sec.~\ref{sect:derivation} and Sec.~\ref{sect:theta_correction}.
For the HvLET effective area satisfying {\tt evtype=L}, a slightly increase is found between 2~GeV and 5~GeV which can be largely explained by the `failed' HET events.
The growth of the LET trigger threshold at the reconstructed energy $\gtrsim 2~\rm GeV$ seems not significant enough to affect the HvLET effective area.

Making MC simulation by adjusting thresholds with time is a choice to achieve the accurate instrumental performance for cosmic-ray study~\citep{Li2023}.
However, due to the anisotropic distribution of both the $\gamma$-ray sky and the exposure map, making such instrumental responses with simulation for $\gamma$-ray study can be quite computationally expensive.
In Sec.~\ref{sect:correction}, we present the parameterized correction to the effective area and the exposure.
We show in Fig.~\ref{fig:exposure_ratio} that the HET and total exposures decrease by $\sim 16\%$ and $\sim 12\%$ on average at 2~GeV, respectively.
The correction only matters for the analyses below 8~GeV.

Finally, we verify the corrections using the Vela pulsar in Sec.~\ref{sect:verify}.
The Vela spectra with the exposure corrections are more consistent with that in the \emph{Fermi}-LAT point source catalog than the one without correction.

%%%%%%%%%%%%%%%%%%%%%%%%%% ACKNOWLEDGEMENT %%%%%%%%%%%%%%%%%%%%%%%%%%%
\begin{acknowledgments}
  We appreciate the anonymous referee for the valuable suggestions and thank Prof. Yi-Zhong~Fan for the help.
  The DAMPE mission was funded by the strategic priority science and technology projects in space science of Chinese Academy of Sciences.
  The data analysis is supported in part by
  %% list of funds
  the National Key Research and Development Program of China (No.~2022YFF0503301),
  the National Natural Science Foundation of China (No.~12003074),
  the Strategic Priority Program on Space Science of Chinese Academy of Sciences (No.~E02212A02S),
  the Project for Young Scientists in Basic Research of the Chinese Academy of Sciences (No.~YSBR-092),
  the Youth Innovation Promotion Association CAS,
  and
  the Entrepreneurship and Innovation Program of Jiangsu Province.

  \software{{\sc NumPy}~\citep{numpy2020}, {\sc SciPy}~\citep{scipy2020}, {\sc Matplotlib}~\citep{matplotlib2007}, {\sc Astropy}~\citep{astropy2018}, {\sc iminuit}~\citep{iminuit2020,Minuit1975}, {\sc DmpST}~\citep{Duan2019}, {\sc Geant4}~\citep{GEANT4:2002zbu,Allison:2006ve,GEANT4}.}
\end{acknowledgments}

%%% appendix
\appendix
\section{The time evolution of the simulated effective area}
We simulate the HET photon data with the trigger thresholds at 0 days, 1500 days, and 3000 days after launch.
The simulation is performed using the {\sc Geant4}~\citep{GEANT4:2002zbu,Allison:2006ve,GEANT4} based on the DAMPE geometric model of the payload and the satellite.
After the photons generation, transportation simulation, digitization, and reconstructions, the simulated photon data are obtained.
To account for the evolution of the trigger thresholds, we change the energy thresholds of the top four BGO layers using the increase rates in~\citet{Li2023} during the digitization process.
More details on the DAMPE simulation can be found in~\citet{Jiang:2020nph,Jiang:2021cit}.

\begin{figure*}[!htb]
  \centering
  \includegraphics[width=\columnwidth]{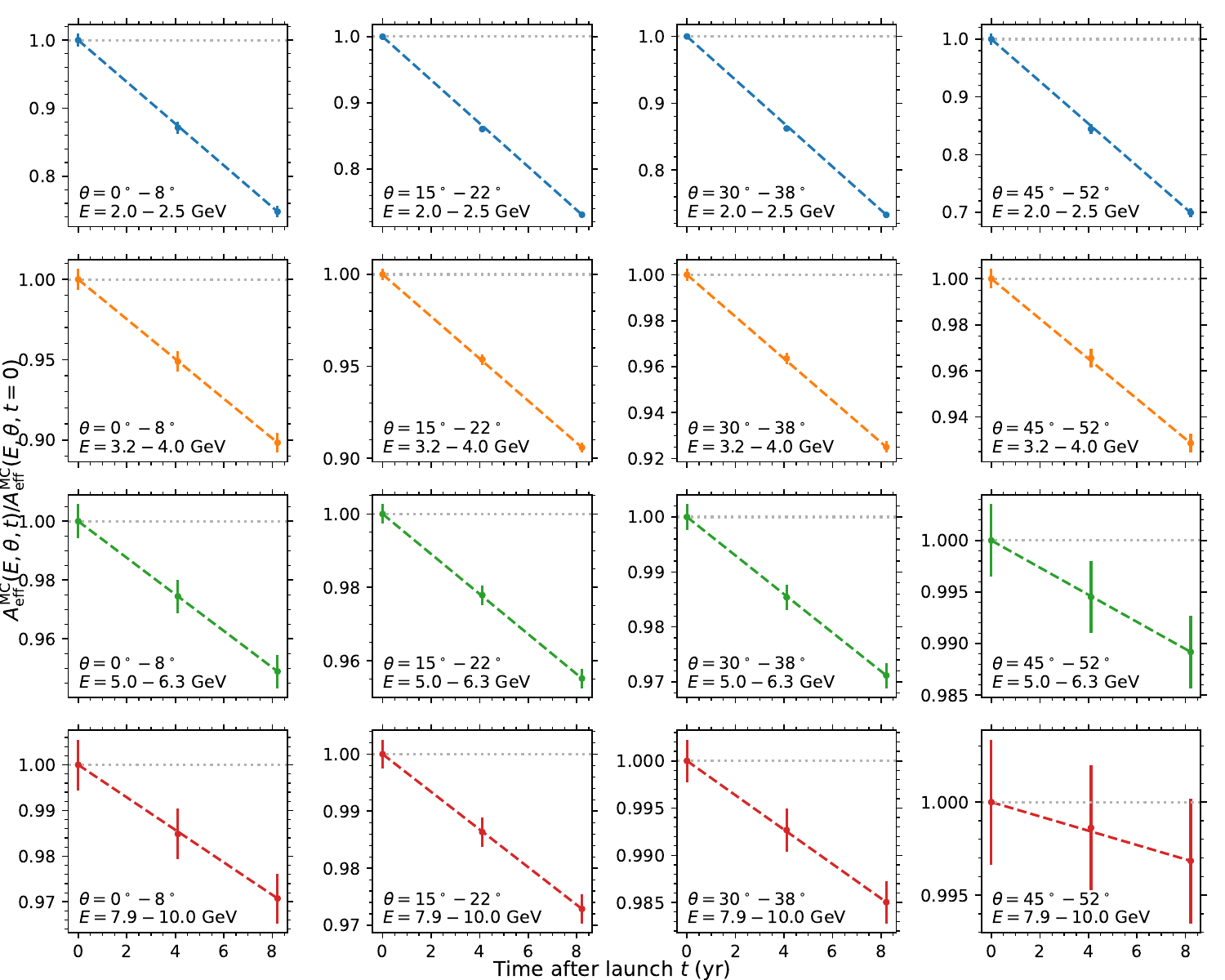}
  \caption{
    The time evolution of the simulated HET effective area within various bins of primary energies and inclination angles.
    The points and error bars correspond to the values and uncertainties of the relative effective area derived from the MC simulation.
    The dashed line in each sub-figure is the linear evolution model.
  }\label{fig:het_simuaff_evo}
\end{figure*}

The simulated data are then partitioned into various energy and inclination-angle bins to make the MC effective areas at different time points.
We divide the effective areas by the corresponding values at $t=0$ whose time evolution is shown in Fig.~\ref{fig:het_simuaff_evo}.
The change of the MC effective area can also be well modeled by the linear function (dashed line).
More details can be found in Sec.~\ref{sect:theta_correction}.

%%%%%%%%%%%%%%%%%%%%%%%%%%% BIBLIOGRAPHY %%%%%%%%%%%%%%%%%%%%%%%%%%%%%
% \bibliography{mybibs} % bibtex file name (w/o suffix)
% \bibliographystyle{aasjournal}

\end{document}